## About the directional properties of Solar Spicules from Hough Transform analysis

E. Tavabi

[a] Physics Department, Payame Noor University, 19395-3697 Tehran, I, R. of Iran

E-mail addresses: tavabi@iap.fr (E. Tavabi)

## ABSTRACT

Spicules are intermittently rising above the surface of the Sun eruptions; EUV jets are now also reported in immediately above layers. The variation of spicule orientation with respect to the solar latitude, presumably reflecting the confinement and the focusing of ejecta by the surrounding global coronal magnetic field, is an important parameter to understand their dynamical properties.A wealth of high resolution images of limb spicules are made available in H CaII emission from the SOT/Hinode mission. Furthermore, the Hough transform is applied to the resulting images for making a statistical analysis of spicule orientations in different regions around the solar limb, from the pole to the equator. Spicules are visible in a radial direction in the polar regions with a tilt angle ($<20^0$). The tilt angle is even reduced to 10 degrees inside the coronal hole with open magnetic field lines and at the lower latitude the tilt angle reaches values in excess of 50 degree. Usually, around an active region they show a wide range of apparent angle variations from -60 to +60 degrees, which is in close resemblance to the rosettes made of dark mottles and fibrils in projection on the solar disk

## 1 INTRODUCTION

The origin of the upper solar atmosphere **heating** and the **fast wind mass loss** from coronal holes (CH) seems directly related to the dynamic behavior of the solar spicules or better, of regions directly above having a more extended structure with giant or macro-spicules and spikes (Koutchmy & Loucif 1991, Koutchmy et al. 1998) as again recently introduced by many authors, for ex. by Sterling et al. (2015) considering the SXR jet counterparts of these important dynamical phenomena. Filippov & Koutchmy (2000), Lorrain & Koutchmy (1998) and many others directly relate spicule orientation to the ambient coronal magnetic field. Auchère et al. (1998), and Filippov, Koutchmy and Vilinga (2007) claimed that it could be related to a chromosphere prolateness effect well observed in "cool" collisional chromospheric lines like Hα and HCaII lines see Johannesson & Zirin (1996) but not observed in radiatively excited HeI lines, see Filippov & Koutchmy (2000). Following a suggestion by Suematsu & Takeuchi (1991), De Pontieu et al. (2004) claimed a relation between inclinations of magnetic flux tubes and the tunneling of low-frequency photospheric P-modes which are more than sufficient in carrying energy fluxes into the corona and transition region. The mechanism could work for more inclined field lines than what is seen in CH with open toward the corona field lines. Classically, it is admitted that less than one percent of the spicule mass is indeed transported towards the corona and this is enough to compensate for solar fast wind mass losses located in coronal holes (see Athay 1986; Wilhelm et al. 2011), where the spicules are more vertical and taller than the quieter spicules outside CH. Giant and macro spicules are believed to sometimes give an EUV and even a SXR (Soft X-Ray) jet, see for *ex.* Koutchmy et al. (1998) for polar CH, which could be explained in case of a release of sufficient energy coming from the free energy of the coronal magnetic field see Filippov et al. (2013), Moore et al. (2015) for processing occurring near a magnetic null point, including reconnection events. Fig. 1 illustrates the complexity of the chromospheric and of the transition region (TR) shells around a polar region with different components simultaneously observed using the best available today EUV images from AIA of the SDO mission: i/ the forest of 304 HeII macro- spicules shown in the upper display as a thick extended towards the corona shell; ii/ the spicule thin fringe seen in absorption at the limb in the 194 image (lower display) with many new features seen in emission crossing the limb and appearing above, including short and long jets and small enhancements. The scale is the same for both displays; disk features can be correlated to illustrate the difficulties to simultaneously consider the intermixed structures of different temperature, a difficulty that will not be resolved in our paper.

Several authors have listed the physical parameters of low excitation emission lines spicules (Mouradian 1965; Beckers 1968; Sterling 2000; Pasachoff et al. 2009, Tavabi et al. 2011a and others) based on off-disk measurements of projected linear structures at a known true angle with respect to the solar local vertical (Mosher & Pope, 1977; Koutchmy & Loucif 1991). They found that the most common inclination is in the range of 20-45 degrees and the average true inclination is about 36 degrees for measurements made below the $70^0$ latitude. These are somewhat broader than the distributions found by Beckers (1968) who gives a variation from the vertical of 20 degrees at a $60^0$ latitude. Van de Hulst (1953) from eclipse observations could not specify the average orientation. Heristchi & Mouradian (1992) found the average inclination around the axis of symmetry is 29 degrees and the average spicule tends to lean toward the equator. More recently, Pasachoff et al. (2009) recorded measurements near the solar minimum activity and found a tilt of 27 degrees with a 5 degrees dispersion, while Heristchi and Mouradian's measurements were taken in 1970, at solar maximum.

The orientation of spicules is a valuable parameter in absence of direct magnetic field measurements of sufficient resolution in this chromospheric region, because it is presumably determined by the flow of plasma which should occur along the magnetic field lines especially where the solar magnetic field pressure dominates the gaz pressure. Of course, all these measurements suffered from overlapping effect of spicules seen along each line of sight, the effect of which will be more important when we look near the solar limb, see Tavabi, et al. (2011b) for a model investigation of this effect. In case of macrospicules as well imaged by AIA of the SDO mission using the 304 filter recording the emissions of the HeII resonance line, an additional effect arises due to the optical thickness of the line, especially on disk and also above the limb in the inner chromospheric shell, see Fig. 1. We will not discuss in great details the case due to the absence of obvious interpretation when these TR emissions are compared with the most typical emissions illustrated in the lower part of Fig. 1.

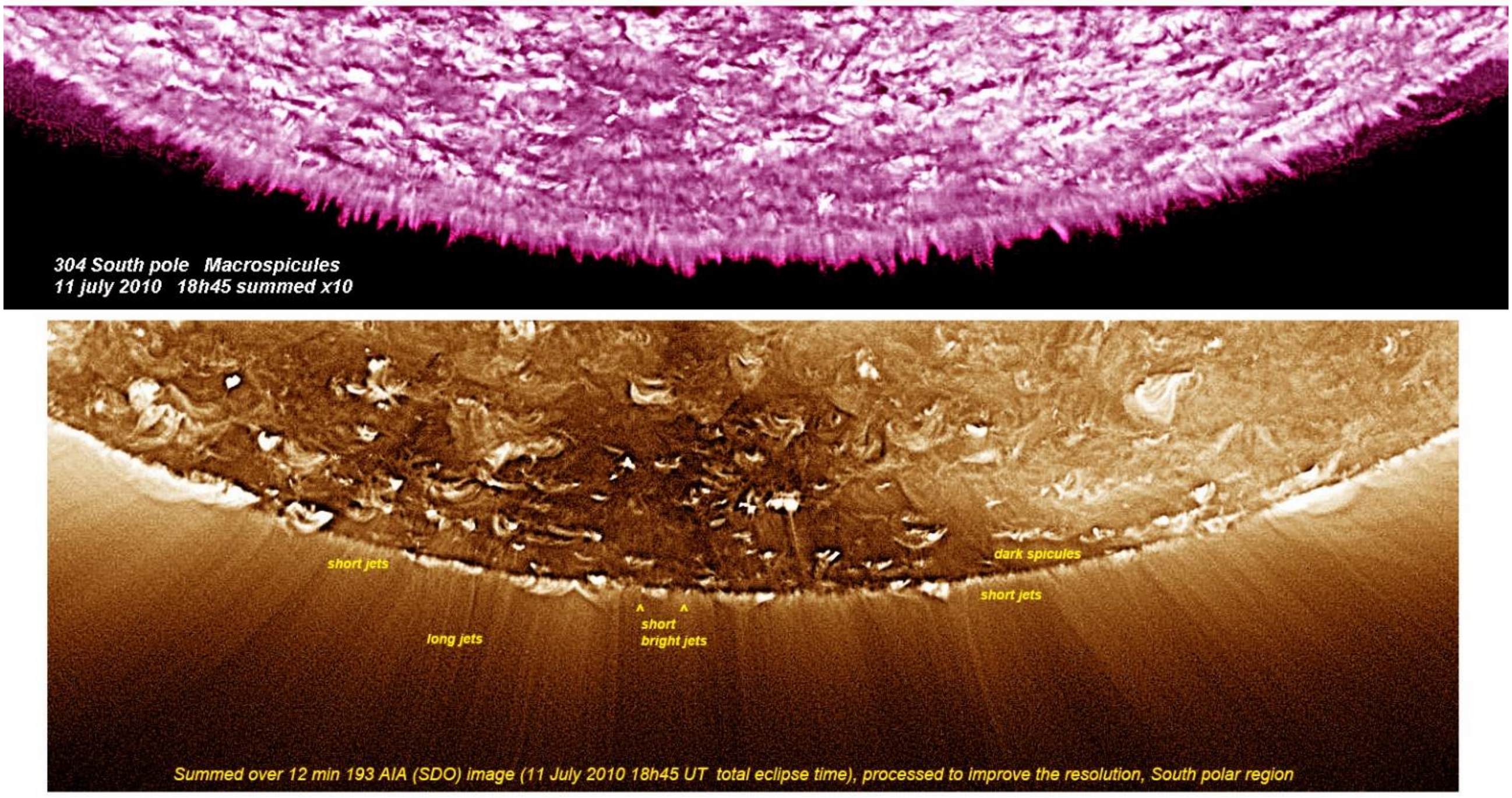

**Figure 1.** The macrospicules of the South polar region observed at the time of the total solar eclipse of 2010 to illustrate the complexity of the TR shell made of 304 HeII emission jets and its coronal counterpart illustrated in the lower image. 10 Images summed around 18h45 are used to improve the signal/noise ratio and a log display was used with subsequent image processing to improve the contrast of the limb features.

The main purpose of this paper is to automatically and objectively determine the apparent tilt angle of spicules using the best available highly processed observations free of seeing effects from the Solar Optical Telescope (SOT) limb imaging experiment using a HCaII line filter see Tsuneta et al. (2008) and Suematsu et al. (2008), onboard Hinode mission (Kosugi et al. 2007). A technique for the automatic detection of limb spicules has been developed and statistical measurements were conducted for determining the tilt angle for spicules at different heliocentric angles.

## 2 OBSERVATIONS AND DATA REDUCTION

We selected five sequences of solar limb observations made at different positions of the limb using the broad-band filter instrument (BFI) of the SOT of the Hinode mission (table 1). We use series of image sequences obtained in the Ca II H chromospheric but "cool" emission line (low excitation line); the wavelength pass-band is centered at 398.86 nm with a FWHM of 0.3 nm and a cadence of 20 seconds is used (with an exposure time of 0.5 s) giving a spatial full resolution of the SOT- Hinode limited by the diffraction at 0¨16 or 120km on the Sun (Kosugi et al. 2007; Goodarzi et al. 2015), a 0¨0541 pixel size scale is used.

The size of each image used here is 1024×512 $pixels^2$ (Hinode read out only the central pixels of the larger detector to keep the high cadence within the telemetry restrictions) thus covering an area of (FOV) 111¨×56¨.

We used the SOT routine program "fg_prep" to reduce the image spikes and jitter effect and to align the time series (Shimizu et al. 2008). The time series show a slow pointing drift, with an average speed less than 0¨015 /min toward the north as illustrated by the very slow solar limb motion.

A superior spatial image processing for thread-like features is obtained using the so- called mad-max algorithm (Koutchmy & Koutchmy 1989; Tavabi et al. 2013), see Fig. 1 and top panels of Figs. 3 and 4 for ex. Table 1 illustrates information on all positions and dates (first and second columns) which are used in this paper.

Fig. 2 shows an example of these processing made to avoid artifact effects due to compression or calibration. For the detection and tracing of spicules in 2D we apply and develop a method with the following steps: (i) To increase the visibility of spicules, a radial logarithmic scale is applied; (ii) To enhance linear features, the Madmax operator is used (Koutchmy & Koutchmy, 1989, Tavabi et al. 2013). The aim of the Madmax operator is to trace the bright hair-like features in solar ultimate observations polluted by a noise of different origins and improve the resolution by non- linearly filtering the data. This popular spatial operator uses the second derivative in an optimally selected direction for which its absolute value has a maximum value (Fig. 2). We use the raw data, but still some artifact phenomena related to pixels size and the transformation methods are seen; note also the CCD read-out defect near the center of the frames in Fig. 3. Next, we consider the spicule apparent inclinations using the Hough transform (Hough 1962).

This section now describes how to use the Hough transform functions to detect lines in an image. This transformation method is a feature extraction technique that finds imperfect instances of objects within a certain class of shapes. The simplest case of Hough transform is the linear transform for detecting straight lines. The Hough transform creates a so- called accumulator matrix. The $(r,\theta)$ pair represents the location of a cell in the accumulator matrix. Every valid (logical true) pixel of the input binary image ($x_i$ , $y_j$) represented by (R,C) produces a r value for all $\theta$ values. The block quantifies the r values to the nearest number in the r vector. The r vector depends on the size of the input image and the user-specified r resolution. The block increments a counter (initially set to zero) in those accumulator array cells represented by $(r,\theta)$ pairs found for each pixel, the straight line can be described as $y_j = Rx_i + C$. This process validates the point (R,C) to be on the line defined by $(r,\theta)$. The block repeats this process for each logical true pixel in the image. The Hough outputs the resulting accumulator matrix. If the curves corresponding to two points are superimposed, the location (in the *Hough space*) where they cross corresponds to a line (in the original image space) that passes through both points. More generally, a set of points that form a straight line will produce sinusoids which cross at the parameters for that line (Shapiro, Linda & Stockman 2001).

The classical Hough transform concerned the identification of lines in images, but nowadays the use of the Hough transform has been extended to identifying positions of arbitrary shapes, most commonly loops and ellipses. Thus, it is a sufficient method to find all linear structures. To qualify them as a spicule, we assume they must be longer than 2" (~20 pixels for SOT images).

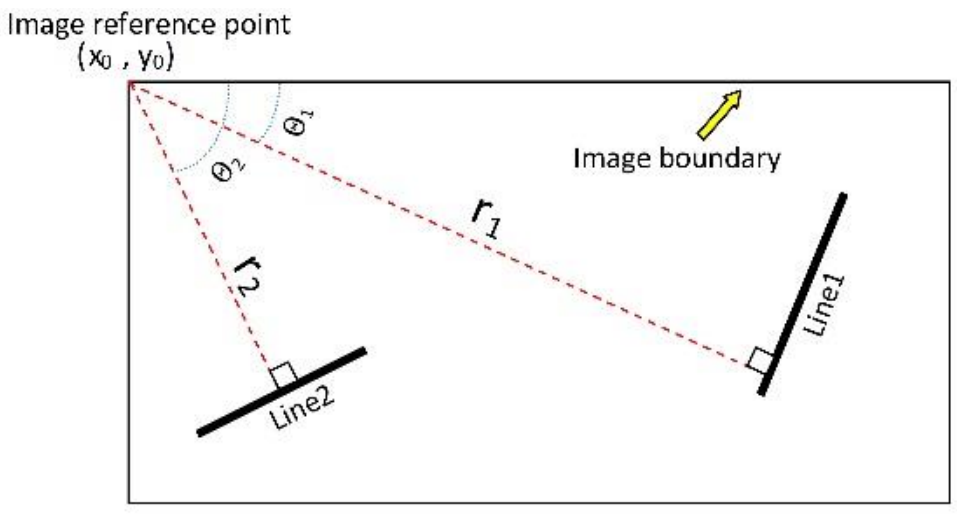

**Figure 2**. At top: negative and processed image taken on 2011-06-17; at Middle: result of line tracing using the Hough transform. The vertical dark line at middle is an artifact due to the CCD readout system, and at bottom: shows the image coordinates.

Figs. 3 and 4 show a result of line segmentation using the Hough transform. With this transformation we can trace more than 70 percent of spicules, seen visually when evaluated using mad-maxed image and the Hough transformed data. In Hough transformation this rate could be increased by changing the segmentation value, to qualify features as spicules. Spicules were defined as bright and straight features of more than 2 arcsec (about 20 pixels) length and of at least 4 pixels width. Using these values we obtained a typical level of accuracy of about 70 percent, although in several cases, at crossing points, such accuracy could not be achieved. The average value for spicule length is about 5,000 to 10,000 km and they have a wide range of widths (in close agreement with Tavabi et al. 2011a).

The tilt angle distribution of the fine H Ca II spicules is the same in both directions left and right from the normal direction with an absolute value of about 50 degrees. This behavior is shown by the spicule tracing using the Hough transform in Fig. 4, and the matrix of the Hough transform clearly showing a statistically equal tilt angle (red color in panel b of Fig. 5).

**3 RESULTS**

We investigated in more details the apparent inclination of spicules and found a statistically average value for different locations around the solar limb. The results are outlined in table 1 including four different locations observed at the minimum of activity; the Hough matrix elements are plotted in Fig. 5.

Furthermore, we found a significant difference between a polar region of the quiet Sun (2008- 2009) Fig. 5 and for the coronal hole region of the active Sun of 2011 (Fig. 6), when the solar B angle had a large value making the coronal hole layers visible (Fig. 3). In this figure, spicules are mostly taller and radial; in addition, some small loops were also seen there.

**4 DISCUSSION**

Our results are indeed in general agreement with the description given in Pasachoff et al. (2009) and Heristchi & Mouradian (1992) from a more morphological method type analysis. The quiet Sun spicules in the lower latitude were oriented in a direction toward the solar equator. This behavior had been reported by Heristchi & Mouradian (1992) and for giant spicules the tendancy was reported in the much earlier seminal paper of Roberts (1945) and later by Koutchmy & Loucif (1991). But Pasachoff et al. (2009) could not clearly find such correlation. In our study, we found that this effect depends on the level of activity of the nearby region for ex. at North East in Fig. 6. A large difference was seen between the two opposite directions as a result of the presence of active regions. To confirm this effect, we would need a larger field of view. The Hough transform matrix also gives us a rough estimate of the number of detected spicules (or straight lines) over the image (table 1, column 3 and the corresponding color bar for each plot).

The solar chromosphere above 1.5 Mm or even less is not a spherically stratified atmosphere as assumed in classical hydrostatic atmospheric models see for ex. Bazin & Koutchmy (2013). The upper edge of the chromosphere seen at moderate resolution in strong chromospheric emission lines is rather "blurred" as it consists of the mixture of a large number of jet-like dynamic spicules and of coronal plasma between them. As far as we know, the Chromospheric layer is mainly filled with spicules/jets and the thickness of the chromosphere shows a wide range of variations (from pole to equator and quiet to active Sun during the solar cycle). This has been reported by several authors (Auchère et al. 1998; Zhang et al. 1998; Filippov & Koutchmy 2000). They suggested that the elevation of the limb in the chromosphere may be caused by the presence of spicules shown in low excitation emission lines, leaving rather open the question of much hotter EUV jets and of macro- spicules, see Fig. 2. Many past observations showed that at the epoch of solar minimum the extension of the chomosphere near the poles is systematically higher than at the equator (Secchi 1877; Roberts 1946; Auchere et al. 1998; Johannesson & Zirin 1996; Filippov & Koutchmy 2000). The amount of prolateness depends of the behavior of spicules and possibly of the inter-spicule matter see Filippov & Koutchmy (2000) and Filippov, Koutchmy, and Vilinga (2007). The difference in the height of polar and equatorial chromospheres arises due to the difference in structure of polar and low latitude magnetic fields. It is well-known that the large-scale magnetic field in the polar regions is mostly open at the sunspot minimum, while in the equatorial region it is mostly closed according to the dominance of the global dipolar and octupolar spherical harmonics (Filippov, 2008). The small-scale structure of the magnetic field in the two regions is likely not being the same. The real fine structure of the polar magnetic field is still elusive. There are strong suspicions that it is very similar to the fine structure of the magnetic field within coronal holes sometimes observed at low latitudes. Here we added the result of analysis of another parameter linked to the direction of the field which is showing that both the extension and the directions are related.

Recent space-borne observations confirmed and substantiated these early suggestions see for ex. De Pontieu et al. (2007), Moore et al. (2015) by considering the jets above, starting with macrospicules seen in 304 HeII line emission see Fig. 1. Even the limb EUV eruptive events of Tavabi, Koutchmy and Golub (2015) could be included in the analysis. Fig. 1 did show that this is a complex question because the relations between the intermixed components of different temperatures is not clear. This question needs in more analysis and we will consider it in a forthcoming paper.

Finally let us note that the polar extension seems consistent with a reduced heat input to the chromosphere in the polar coronal holes where the coronal T° above is known to be lower, compared to the quiet-Sun atmosphere at the equator vs. the active-Sun, without taking into account the most energetic explosive type events that are recorded also in polar regions.

**ACKNOLEDGEMENTS**

We are grateful to the Hinode team for their wonderful observations. Hinode is a Japanese mission developed and launched by ISAS/JAXA, with NAOJ as a domestic partner and NASA, ESA and STFC (UK) as international partners. The image processing OMC software (often called Madmax) written for IDL is easily downloadable from the O. Koutchmy site at UPMC, http://www.ann.jussieu.fr/~koutchmy/debruitage/madmax.pro. We thank C. Bazin for helping us with the preparation of Fig. 1, and A. Ajabshirzadeh, B. Filippov and L. Golub for meaningful discussions.